\documentclass[12pt,preprint]{emulateapj}
\shorttitle{High-mass protostar NGC\,7538\,S$_{\rm A}$}
\shortauthors{Wright et al.}
\begin{document}

\def\arcmin{{$^{\prime}$}}
\def\arcsec{{$^{\prime\prime}$}}
\def\ptsec{$''\mskip-7.6mu.\,$}
\def\psec{$^s\mskip-7.6mu.\,$}
\def\Msun{\,{\rm M$_{\odot}$}}
\def\Lsun{\,{\rm L$_{\odot}$}}
\def\ltsim{$\stackrel{<}{\sim}$}
\def\gtsim{$\stackrel{>}{\sim}$}
\def\jtra#1#2{${\rm J}\!\!=\!\!{#1}\!\!\to\!\!{#2}$}

%** added
\def\degr{$^{\circ}$}

\title{Observations of  a high-mass protostar in NGC\,7538\,S}

\author{ Melvyn Wright}
\affil{Radio Astronomy Laboratory, University of California, Berkeley
601 Campbell Hall, Berkeley, CA 94720, USA}

\author{ Jun-Hui Zhao}
\affil{Harvard-Smithsonian Center for Astrophysics, 60
  Garden Street, Cambridge, MA 02138, USA}
\author{G\"oran Sandell}
\affil{NASA Ames Research Center, SOFIA-USRA, Mail Stop 211-3,
Building N211, Rm. 249, P.O. Box 1, Moffett Field, CA 94035-0001, USA}

\author{Stuartt Corder}
\affil{National Radio Astronomy Observatory, 520 Edgemont Road, Charlottesville, VA, 22903, USA}
%\altaffiltext{1}{Jansky Fellow, NRAO}

\author{W. M. Goss}
\affil{National Radio Astronomy Observatory, P.O. Box O, Socorro, NM 87801, USA}

\author{Lei Zhu\altaffilmark{2}}
\affil{Department of Astronomy, Peking University, Beijing 100871, China}
\altaffiltext{2}{Predoctoral Fellow, Smithsonian Astrophysical Observatory} 

\begin{abstract}

We present high angular resolution continuum observations of the high-mass
protostar NGC\,7538S with BIMA and CARMA at 3 and 1.4 mm, VLA observations
at 1.3, 2, 3.5 and 6 cm, and archive IRAC observations from the {\it Spitzer} Space
Observatory, which detect the star at 4.5, 5.8, and 8 $\mu$m.  The star looks rather unremarkable in the
mid-IR. The excellent positional agreement of the IRAC source with the VLA free-free
emission, the OH, CH$_3$OH, H$_2$O masers,  and the dust continuum
confirms that this is the most luminous object in the NGC\,7538\,S core.
The continuum emission at millimeter wavelengths is dominated by dust
emission from the
dense cold cloud core surrounding the protostar. Including all array configurations,
the emission is dominated by an elliptical source
with a size of $\sim$ 8$^{\prime\prime}$ $\times$ 3$^{\prime\prime}$. If we filter out the extended 
emission we find three compact mm-sources inside the elliptical core. 
The strongest one, $S_A$,
coincides with the VLA/IRAC source and resolves into a double source at 1.4 mm, where we 
have sub-arcsecond resolution.
The measured spectral index, $\alpha$, between 3 and 1.4 mm is $\sim$ 2.3,
and steeper at longer wavelengths,  suggesting a low dust emissivity or
that the dust is optically thick. We argue that the dust in these accretion disks 
is optically thick and estimate a mass  of an accretion disk or infalling envelope surrounding S$_A$ to be $\sim$ 60M$_{\odot}$ 

\end{abstract}

\keywords{ISM: clouds -- (stars:) circumstellar matter -- stars:
formation -- stars: pre-main sequence -- submillimeter}

\section{Introduction}
\label{Intro}

The giant molecular cloud southeast of the \ion{H}{2} region NGC\,7538,
at a distance of  2.65 $\pm$ 0.12 kpc \citep{Moscadelli09}, is a 
well known site of high-mass star formation \citep{Werner79,Qiu11}.  
The early work by \citet{Werner79} showed that star formation occurs 
in three activity centers, each of which maybe in a different 
evolutionary stage. The IRS\,1 - 3 region dominates the luminosity 
(L$_{\rm bol} \sim$ 2 10$^5$ \Lsun{})  and is the closest to the 
\ion{H}{2} region. Each of the three infrared sources coincides with 
ultracompact \ion{H}{2} regions. NGC\,7538 harbors
several young massive stars, of which at least three (IRS\,1, IRS\,9 and
NGC\,7538\,S) drive powerful molecular outflows and appear to be
surrounded by accretion disks. 

The massive, $\sim$25 \Msun, young star
IRS\,1 appears to be the most massive member of a young embedded cluster
\citep{Qiu11}. It powers a collimated thermal jet \citep{Sandell09} and
drives an extended molecular outflow \citep{Corder08,Qiu11}. IRS\,1 is
still heavily accreting \citep{Corder08, Qiu11}, i.e.,  it must be
surrounded by an accretion disk. However, no one has yet managed to
directly detect and image the disk.

NGC\,7538\,S, $\sim$ 85\arcsec\ 
south of IRS\,1  is much colder, has lower luminosity (L$_{\rm bol} \sim$ 
1.3 10$^4$ \Lsun{}) and coincides with maser emission from OH, CH$_3$OH,  and 
H$_2$O, but not any near-IR source, while the third region, IRS\,9 
(L$_{\rm bol} \sim$ 2 10$^4$\Lsun{}), was found to be associated with 
a bright near- and mid-infrared source. Submillimeter continuum imaging 
at high angular resolution starts to resolve the molecular cloud into 
a lumpy filamentary structure connecting the individual cores \citep{Sandell04,Reid05}. 
IRS 9 illuminates a prominent
reflection nebulosity, has only weak free-free emission and drives an
extreme high-velocity outflow seen almost pole on \citep{Sandell05}.
\citet{Sandell05} found several outflows in the vicinity of IRS\,9,
suggesting that there is also  a young cluster surrounding IRS\,9.

\begin{figure*}[t]
\centering
\includegraphics[width=150mm,angle=0]{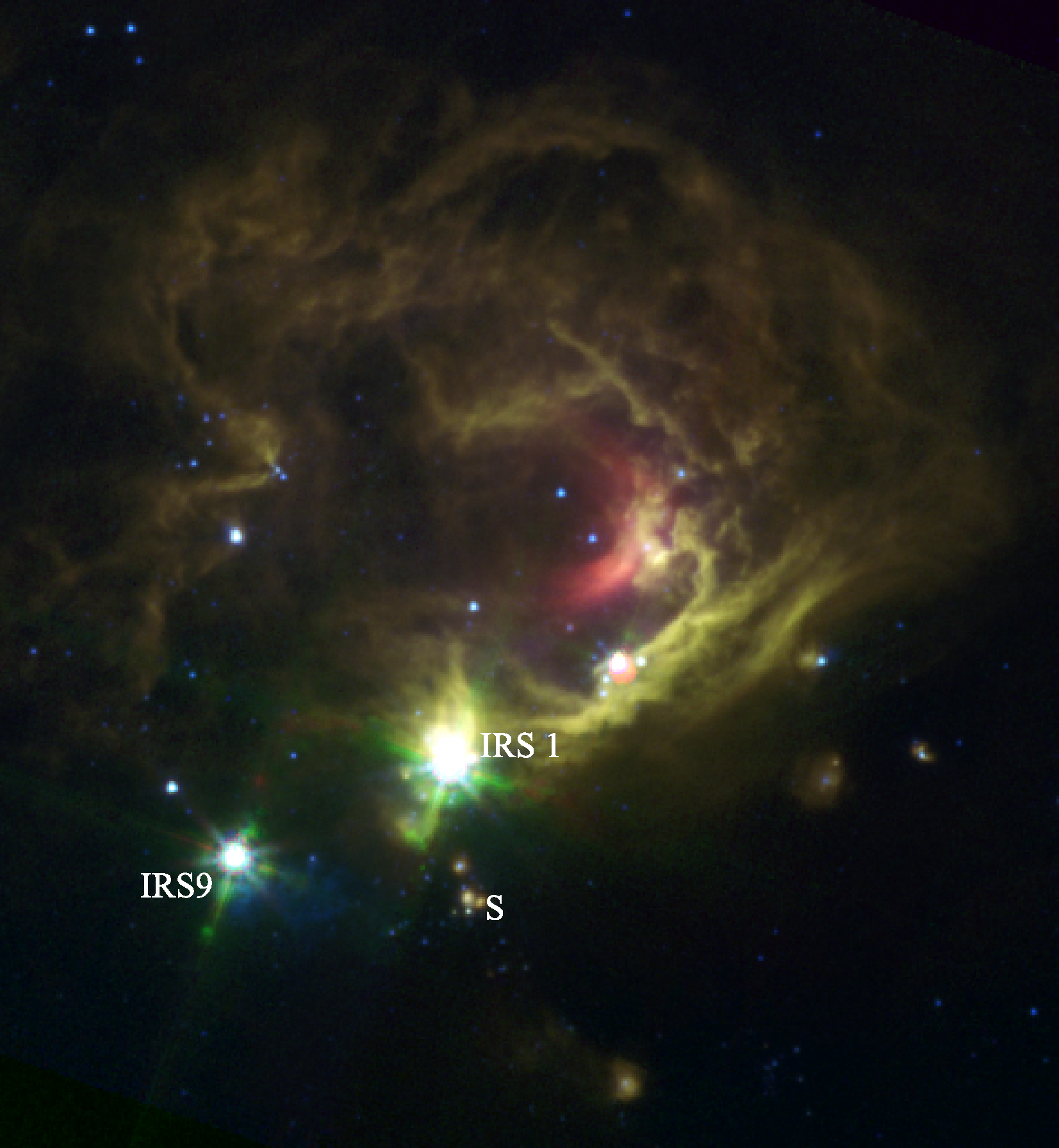}
\caption{This false color {\it Spitzer} IRAC image shows the location 
of IRS\,1, IRS\,9 and S relative to the NGC\,7538  \ion{H}{2} region. 
The 4.5 $\mu$m filter is coded blue, the 5.8 $\mu$m filter is 
coded green and the 8.0 $\mu$m filter is coded red. The boundaries 
of the \ion{H}{2} region stand out due to strong PAH emission from 
the surrounding molecular cloud. IRS\,1, IRS\,9 and IRS\,4$\sim$ 
80\arcsec\ northwest of IRS\,1 are all saturated in the IRAC images.}
\label{fig-color}
\end{figure*}

Figure \ref{fig-color} shows a large IRAC color image of NGC\,7538 
with the three major star forming regions labeled as IRS\,1, IRS\,9 
and S, respectively. NGC7538\,S is a faint, barely visible source 
at the outskirt of a small cluster of IR sources. NGC\,7538\,S is 
not seen at 3.6 $\mu$m and is not even particularly red due to 
the strong absorption from the surrounding cloud (Section \ref{IRAC-results}).

In this paper we focus on NGC\,7538\,S, which is the youngest high-mass
forming core in NGC\,7538 and located $\sim$ 80\arcsec\ south of IRS1\citep{Sandell04}. 
In this core there is a massive young star, which coincides with an OH 1665 MHz maser.
\citet{Sandell03} observed the core
with BIMA in 4 mm continuum and in HCO$^+$ and H$^{13}$CN J = $1 \to 0$
with 3\farcs7-resolution and found that it contained a high mass
protostar surrounded by a large rotating circumstellar accretion disk
driving a very compact molecular outflow. Follow-up observations by
\citet{Sandell10} confirm that the star is surrounded by a rotating
accretion disk, smaller than originally thought, and that it may have fragmented into several protostars.

We present new BIMA  and  CARMA observations from  72~GHz to  220
GHz of the NGC\,7538\,S region with spatial resolutions up to $\sim$1\arcsec. 
These observations are combined with observations at 4.9, 8.5,
15 and 22.5 GHz from the VLA, and archive IRAC images (Mid-IR)  from the
{\it Spitzer} Space Telescope.

% FIGURE 1

% OBSERVATIONS - BIMA
\section{Observations}
\subsection{BIMA Observations}

Observations were obtained with the 9-antenna BIMA array between 2002
and 2004 in the B, C, and D array configurations at 72 - 75, 86 - 89,
110, and 217 - 220 GHz. A summary of all the observations is given in
Table \ref{tbl-1}, which also lists the angular resolution and the
rms-noise level achieved at each frequency. Most of these observations
were designed to study the chemistry and kinematics of the accretion disk,
its outflow and the surrounding cold, dense cloud core. However, in each
case we also included one or several 100 MHz wide bands selected to be
free of line emission, thereby enabling us to measure the true continuum
emission from the disk and the surrounding cloud. Table \ref{tbl-1}
therefore also lists the main molecules and isotopes targeted for each
frequency setting.

At 72 - 75 GHz ($\lambda$ 4.1~mm) we combined the line--free data
into a multifrequency synthesis (MFS) image. This image has less 
sensitivity than the rest of our continuum images, partly because
not all receivers could be tuned to this frequency. Because of the low
frequency the synthesized beam is also larger than at other frequencies
(c.f. Table \ref{tbl-1}). At 89~GHz ($\lambda$ 3.4~mm), we combined 
additional data obtained in the B, C and D array configurations with 
the data reported by \citep{Sandell03} to produce deep high fidelity 
continuum images and spectral line images. These observations 
use the same correlator setup and were reduced as described by  
\citet{Sandell03}.

The 110~GHz ($\lambda$ 2.7~mm) observations were the only ones
completely dedicated to continuum observations. Here we made an MFS 
image from observations in the B \& C array configurations using 16
100-MHz bands spanning over 3 GHz in the upper and lower sidebands 
of the 1st LO. At 220~GHz ($\lambda$ 1.4~mm) the only line-free bands
were three 50 MHz windows in the lower sideband, but we had more tracks 
than at 110 GHz, and we were therefore still able to reach an acceptable 
noise level.

The data were reduced and imaged in a standard way using MIRIAD software
\citep{Sault95}. The quasar 0102$+$584 was used as phase calibrator and
Mars and 3C\,84 for flux and passband calibration.  The uncertainty in
the absolute amplitude scale is $\sim$ 20\%, but we were able to improve
the calibration by applying coherence corrections, see Appendix I. 
As a check on the flux density scale  we also imaged
IRS\,1 and IRS\,2,  which are at the edge of the $\sim$ 2 arcmin primary
beam in the 3 mm band.
In this paper we discuss only the continuum data, which allow us to derive
an estimate of the mass and the physical properties of the rotating disk
surrounding the high--mass protostar NGC\,7538\,S. The spectral line
observations and the physical properties of the HII regions associated 
with IRS\,1, IRS\,2 and IRS\,3 will be discussed in Paper II.
 
% OBSERVATIONS - CARMA
\subsection{CARMA Observations}

Observations were obtained with the 15-antenna CARMA telescope in 2009
July - August in the  C,  D, and E  antenna configurations at 220 GHz,
and in the B-configuration in 2009 December and 2010 January at 89, 
110, and 220 GHz. The CARMA observations used two 500 MHz bands in 
upper and lower sidebands of LO1 for a total continuum bandwidth of 
2 GHz. The data were reduced and imaged in a standard way using MIRIAD 
software. 

In these observations we centered the antenna pointing on IRS1 and 
NGC\,7538\,S to minimize the primary beam errors. Previous CARMA mosaic 
observations of NGC\,7538  were designed to map the large scale molecular 
outflows; pointings were not centered on IRS1 or NGC\,7538\,S, and 
uncertainties in the primary beam patterns have prevented accurate 
deconvolution of the sidelobes of IRS1 from NGC\,7538\,S. 
Optical offset pointing was used to ensure accurate pointing of the CARMA antennas \citep{Corder10}.
Rapid switching  between IRS1 and NGC\,7538\,S was used to calibrate NGC\,7538\,S. 
The quasar 0102$+$584 was used as a gain and phase calibrator, and Uranus and MWC\,349 
for flux and passband calibration. The uncertainty in the absolute amplitude 
scale is $\sim$ 20\%. The  strong compact emission from IRS\,1 was used 
to self calibrate the IRS\,1 pointing.

Images were made at 89, 110 and 220 GHz bands by combining the two 500 MHz 
in upper and lower sidebands in a multifrequency synthesis mosiac. The mosaic 
images combine data with the three different primary beams for the 6.1 and 
10.4 m CARMA antennas. The weighted mean observing frequencies from the MFS 
synthesis are 87.8, 111.1 and  222.2 GHz. 

% OBSERVATIONS - VLA
\subsection{VLA Observations}

Our observations with the Very Large Array (VLA) of the National Radio
Astronomy Observatory (NRAO)\footnote{The NRAO is a facility of the
National Science Foundation operated under cooperative agreement by
Associated Universities, Inc.} at 3.6~cm and reduction of unpublished
6~cm archive data were already described in \citet{Sandell05}. Here we
therefore only summarize the observations and the essential parameters
needed to understand and interpret the results. The 3.6~cm observations
were done in the BnA configuration on 2003 October 14. The on-source
integration time was 1.3 hrs and the synthesized beam,  full width half
maximum (FWHM), was 1\ptsec22 $\times$ 0\ptsec47,  (p.a. = $-$2\degr{}).
The rms-noise is  $\sim$ 35 $\mu$Jy~beam$^{-1}$. The peak flux density
on NGC\,7538\,S is  2.3  mJy~beam$^{-1}$.  The 6~cm  observations were
done in the B array on 1989 April 28. The on-source integration time was
2.7 hrs and the synthesized FWHM was 1\ptsec35 $\times$ 1\ptsec09  (p.a.
$-$24\degr{}) with an rms noise $\sim$ 100 $\mu$Jy~beam$^{-1}$.

The 1.3~cm observations were carried out in  the C configuration on
2004 March 19, pointing at NGC\,7538\,S. The integration time was 2.2 hrs. 
These observations were carried out in 4-IF continuum mode, with one 
50 MHz bandwidth pair centered on 22.485 GHz and one 0.195 MHz bandwidth 
pair was centered on the strongest H$_2$O maser component in NGC\,7538\,S, 
see e.g. \citet{Reid97} for a more thorough discussion of this procedure. In
order to determine what maser feature was the strongest at the time
of our observations, we had a short test run in 2004 February, which
showed that the strongest velocity component was centered on V$_{lsr}$
= $-$53.7 km~s$^{-1}$. On March 19, the flux density of this velocity
component was 126 Jy~beam$^{-1}$. The initial calibration of the data
was done using 2322$+$509 as phase calibrator and  3C\,48 as the flux
density calibrator. We then used the strong maser feature as a phase
calibration reference, which considerably increased the signal-to-noise
in our final image. The synthesized beam FWHM for these observations
was 0\ptsec94 $\times$ 0\ptsec78,  (p.a. = $-$10\degr{}).
The rms-noise is $\sim$ 0.22 mJy~beam$^{-1}$. The peak flux density 
in the image is 3.3 mJy~beam$^{-1}$.

We also used a deep 2~cm observation from the
VLA historical archive  (project AM0487) conducted
in the A array configuration on June 30, 1995. The observing frequency
was 14.94 GHz, the bandwidth 50 MHz  and the on source integration time
4.4 hrs. 3C\,84 was used for flux calibration and 2229$+$695 for phase
calibration. The synthesized FWHM is 0\ptsec139 $\times$ 0\ptsec109
and the rms noise $\sim$ 80 $\mu$Jy~beam$^{-1}$. Since the observations
were centered on IRS\,1, the emission from NGC\,7538\,S is attenuated
by about a factor of two by the primary beam  with a peak flux density
of 0.4 mJy~beam$^{-1}$.

% OBSERVATIONS - IRAC
\subsection{ IRAC archive data from Spitzer}
\label{IRAC-data}

We have retrieved and analyzed images of NGC\,7538 taken with the Infrared
Array Camera (IRAC). IRAC is the mid-infrared camera on the Spitzer
Space Telescope with four arrays at 3.6,  4.5, 5.8, and 8.0 $\mu$m.
Each array has 256 $\times$ 256 pixels with the same plate scale,
1\ptsec22 per pixel \citep{Fazio04}.  The observations discussed here
were obtained on December 23, 2003 as part of the GTO program P201
(G. Fazio) and observed in the High Dynamic Range (HDR) mode with a
three point dither for each pointing. In the HDR mode images are taken
with a short (0.6 s) and a long (12 s) integration to capture both
bright and faint sources in each pointing. IRS\,11 \citep{Werner79}
is the brightest source in the NGC\,7538\,S field and it is saturated
in the long integration exposures at 8 $\mu$m, resulting in bandwidth
effect trails\footnote{see the Infrared Array Camera Data Handbook,
http://ssc.spitzer.caltech.edu/irac//dh/iracdatahandbook3.0.pdf for
details.}. At 8 $\mu$m we therefore only use the short exposure images.

We processed both the short (0.6 s) and the long (12 s) integration
basic calibrated data (BCD) IRAC frames in each channel using
the artifact mitigation software developed by Sean Carey and
created mosaics using MOPEX. The saturation
of the 8 $\mu$m images is particularly severe and could not
be corrected for by the available tools. At 8 $\mu$m  we therefore used
only the short integration images to identify sources and derive
photometry. Because the whole NGC\,7538 complex is associated with
extensive nebulosity and strong PDR emission (Figure \ref{fig-color}),
all photometry was derived using PSF fitting with APEX, which is part of the MOPEX reduction package. For sources which APEX failed to detect at one or several wavelengths, we used the APEX user list option to supply the coordinates for the source to successfully derive a PSF fit.

We also retrieved and analyzed Spitzer MIPS images of NGC\,7538. 
NGC\,7538\,S  is not saturated  at 24 micron, but IRS11 is and NGC\,7538\,S  is blended with the
saturated PSF from IRS11, so it is not possible to derive a flux. 
At 70 micron the whole region from IRS1 southward to IRS11 and NGC\,7538\,S
is saturated and contains no information.

% RESULTS - CARMA
\section{Results}
\label{results}

\subsection{CARMA}

\begin{figure}[t]
\centering
\includegraphics[width=68mm,angle=-90]{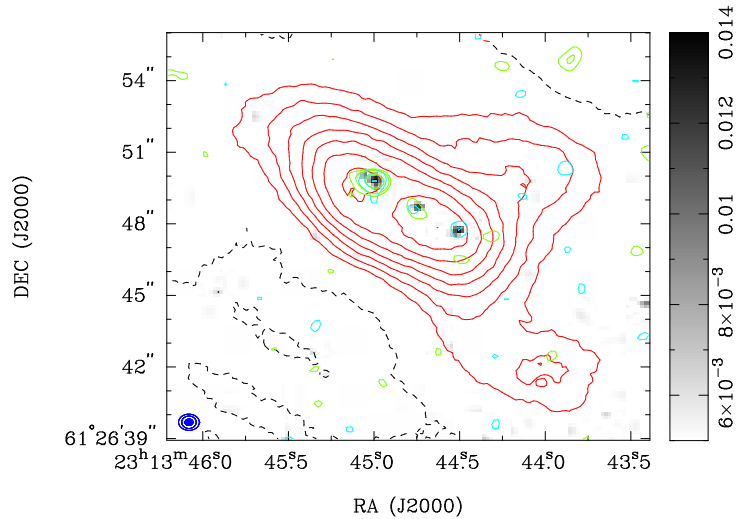}
\caption{
High-resolution images of NGC\,7538\,S made with CARMA in B array configuration
at 222 GHz (gray scale), 111  GHz (cyan contours), and 87 GHz (green contours),
superimposed on the deep image of the NGC\,7538\,S cloud core with CARMA at 224 GHz in the 
B, C, D and E array configurations (red contours).  The beam size (FWHM) for the B-array images at 222, 111 and 87 GHz
are shown at bottom left.
Contour intervals:  13.3 mJy at 222 GHz, 3, 6, 9 mJy at 110, and 2, 3, 6, 9 mJy at 87 GHz
}
\label{224-BCDE-S-sdicm-regrid.pdf}
\end{figure}

Figure~\ref{224-BCDE-S-sdicm-regrid.pdf} shows the deep 222 GHz B, C, D, and E array
image of the NGC\,7538\,S cloud. The emission is dominated by the elliptical core, which at the western side
has an extension to the northwest and a curved filamentary like structure to the southwest.
The northwestern extension has one embedded sub-mm source, the H$_2$O maser IRS\,11S, which we  detected at 111 and 222  GHz, but not at  88 GHz.  In Figure~\ref{224-BCDE-S-sdicm-regrid.pdf} this source is seen as only one cyan contour level. The same source
was also marginally detected by \citet{Corder08}, who labeled it S$_f$.  \citet{Corder08} also found two embedded
protostars in the fainter  core  in the  southwestern filament,  which is labeled as BIMA source South in Table 2. The core is clearly seen in the  deep 222 GHz image,
but we did not have enough sensitivity in the B-array images to recover any embedded sources.  At the 
highest resolution,  0\farcs2, i.e. the 222.2 GHz B-array image (Figure~\ref{224-BCDE-S-sdicm-regrid.pdf} gray scale and Figure 8), the high-mass protostar, S$_A$,  
has been resolved into two components: S$_{A1}$, and S$_{A2}$. The flux densities of the components
were estimated from Gaussian fits. The positions, integrated fluxes, deconvolved 
sizes,  and position angles are given in Table 2.     % ~\ref{tbl-3}.

% FIGURE 2
%(options=mfs,mosaic,double,systemp robust=0.5 )

% RESULTS - VLA

\subsection{VLA}
\label{VLA-results}
\begin{figure}[t]
\centering
\includegraphics[width=85mm, angle=0]{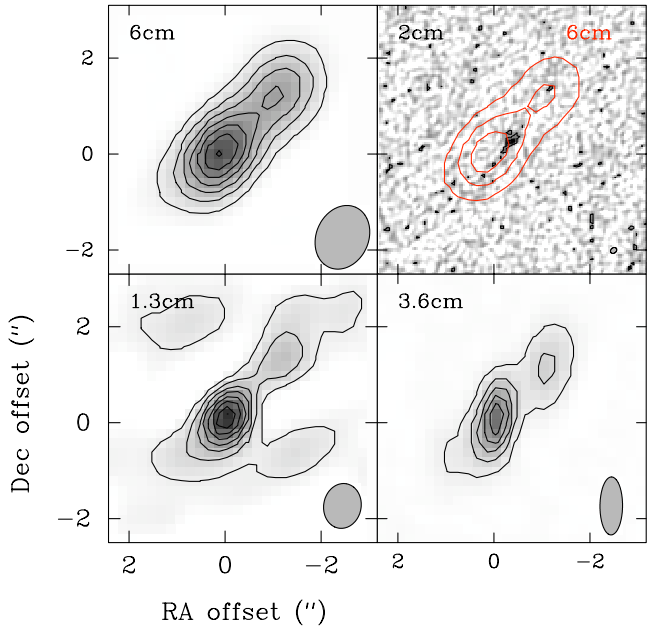}
\caption{
VLA continuum images centered on NGC\,7538\,S at 1.3, 3.6, 2 and
6~cm. The 2~cm image (grey-scale only) had IRS\,1 as pointing center and
has poor SNR because it is in the outskirts of the primary beam and has
very high angular resolution, $\sim$ 0\ptsec1. The integrated intensity,
however, agrees with the flux densities measured at other frequencies and
confirms that the free-free emission originates in a highly collimated
ionized jet. Position offsets are w.r.t. NGC\,7538\,$S_a$
}
\label{fig-vla}
\end{figure}

The VLA observations show a double source at 4.9, 8.5, and 22.5 GHz
(Figure \ref{fig-vla}). The stronger component (a) is centered on
NGC\,7538\,S, while the second component (b) is $\sim$ 1\ptsec5 
to the north west at a position angle of $-$34\degr. Both components
appear elliptical, and Gaussian fits, when the components are sufficiently
resolved (Table 2), show that their major axis is aligned in the same
direction, suggesting that the two components are part of a collimated jet
emanating from NGC\,7538\,S into the blue-shifted molecular outflow lobe
powered by NGC\,7538\,S \citep{Sandell03}. The high angular resolution
image at 14.9 GHz confirms that the main component has the appearance of
a highly collimated jet, although it does not have enough sensitivity
to clearly detect the northwestern component, which in this picture
would be the tip of the jet. Our deepest image, at 8.5 GHz, reveals a
faint extension to the southwest, which we interpret as the counter jet
traveling into the red-shifted outflow. A least squares fit to the sum
of both components yields a spectral index of 0.14, which confirms that
we are looking at wind-ionized free-free emission from a collimated jet,
see e.g.  \citet{Reynolds86}.

%Our observations at 4.9 and 8.5 GHz also cover IRS\,1 - 3. In Table 2 %\ref{tbl-2} 
%we therefore list the positions and flux densities, which we have deduced
%from Gaussian fitting. In Table 2 %\ref{tbl-2} 
%we have also included an unpublished
%BIMA observations at 36.6 GHz (Forster, pers. comm). It should be noted,
%that even though IRS\,1 is usually referred to as an ultracompact or
%hypercompact \ion{H}{2} region, it does not have a spectrum  resembling
%a normal ultra compact \ion{H}{2} region. When IRS\,1 is imaged at
%sub-arcsecond resolution \citep{Campbell84,Gaume95} it shows a double
%lobed  north-south core with a size of $\sim$ 0\ptsec4 with faint
%extended lobes extending about 1\arcsec\ to the north and to the south.
%\citet{Lugo04} have made a compilation of all published flux density
%measurements of IRS\,1 and model the emission as a photo-evaporated
%disk wind. If we use the data presented in \citeauthor{Lugo04} combined
%with our own data and do a least squares fit from 1.6 GHz to 36 GHz,
%we derive a spectral index $\alpha \sim$ 1.0  over the whole frequency
%range, which would exceed the observed flux density at a 100 GHz (see
%Discussion). It is therefore almost certain that the compact core is
%optically thick to $\sim$ 30 GHz or more, and that the diffuse lobes are
%wind-ionized free-free emission from a collimated jet 
%and have a much flatter spectrum, perhaps similar to that of
%NGC\,7538\,S. 

% FIGURE 3

% RESULTS - BIMA
\subsection{BIMA}
\label{BIMA-results}

\begin{figure}[t]
\centering
\includegraphics[width=67mm, angle=-90]{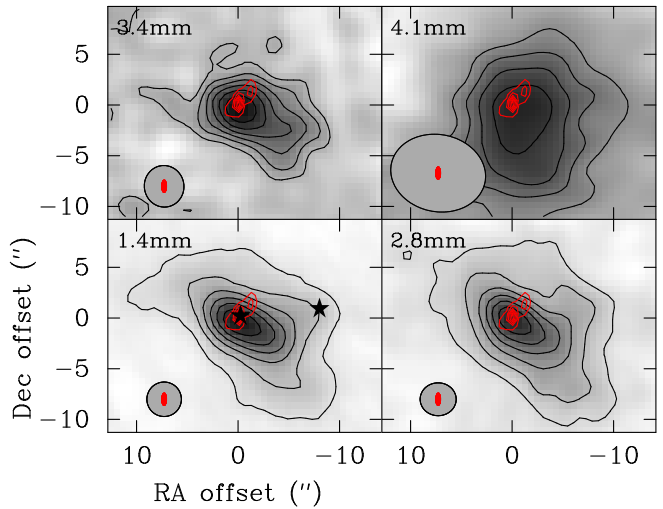}
\caption{
BIMA continuum images of NGC\,7538\,S  at 1.4, 2.8, 3.4, and 4.1 mm
in grayscale overlaid with contours. The red contours on each image
shows the free-free emission at 3.6 cm. On the 1.4 mm image we have
additionally marked the position of the two IRAC mid-IR sources coinciding
with NGC\,7538\,S and the western H$_2$O maser source IRS\,11\,S (Table 3).
%\ref{tbl-2}).  
For each image we have plotted 6 linearly spaced contours
between $\sim$ 3-$\sigma$ rms level and peak flux density. The beam FWHM
is indicated at the lower left of each panel.
Position offsets are w.r.t. NGC\,7538\,$S_a$
}
\label{fig-mmcont}
\end{figure}

Images of NGC\,7538\,S at 74, 89, 109, and 217 GHz are shown in Figure \ref{fig-mmcont}.
The 109 and 217 GHz images appear very similar and agree quite well with the deep
CARMA image at 222 GHz, although the CARMA image recovers more of 
the extended emission in the cloud core  (Fig.~\ref{224-BCDE-S-sdicm-regrid.pdf}). 
Both the 109 and the 217 GHz image  show extended dust
emission with a size of $\sim$ 20\arcsec. The emission is dominated
by an elliptical core  centered $\sim$ 1\arcsec\ -
2\arcsec\ southwest of NGC\,7538\,S with a position angle of $\sim$
50\degr. They also show two secondary peaks, one $\sim$  5\arcsec\
west and 1\arcsec\ north, the other $\sim$ 5\arcsec\ west and 6\arcsec\
south of NGC\,7538\,S. The former coincides within errors with an H$_2$O
maser, which is also detected in all IRAC bands from 3.6 - 8 $\mu$m, see
Section \ref{IRAC-results}. The H$_2$O maser was also detected as source D
in the CARMA B-array observations (Table 2).  The southwestern core has 
no maser or IR counterpart. \citet{Corder08} identified two  3~mm sources in this 
core, but we have not been able to confirm them.
%and is therefore likely to be dense dust condensation in the cloud,
%which may at some point collapse to form a star. At 89 GHz the ``disk''
%appears smaller and more centrally condensed than at 109 and 217 GHz,
%see also Table 2 %\ref{tbl-2}. 
%The two secondary peaks are still marginally
%visible in the 89 GHz image. 
The image at 74 GHz looks different. Here the emission is much fainter than at
the higher frequencies and the spatial resolution is poorer. The
difference in morphology is most likely due to poor image fidelity and sensitivity.

% FIGURE 4

 % DISCUSSION
\section{Discussion}

\subsection{The embedded stellar population - IRAC mid-IR imaging }
\label{IRAC-results}
Table 3 lists the positions of the 14 sources which were  detected at 8
$\mu$m within a 25\arcsec\ radius around NGC\,7538\,S. NGC\,7538\,S  is
detected at 4.5, 5.8, and 8 $\mu$m, but not at 3.6 $\mu$m. There is a
small cluster  of four sources centered on IRS\,11, and another faint
source east of NGC\,7538\,S. One of these sources, $\sim$
5\arcsec\ west of NGC\,7538\,S coincides with an H$_2$O maser and is also
seen as a compact dust emission source in the BIMA continuum images at
109 and 217 GHz.

In Figure 6 we plot all the 14 sources from Table 3 in an IRAC color-color plot.
 Six of the 14 stars in Table 3 have  
 low color indexes ([5.8]-[8.0] $<$ 0.4 and [3.6]-[4.5] $<$ 1.5)
and are most likely reddened field stars. Two of them, J23134708+6127090 and J23134160+6126427, however,  have  [5.8]-[8.0] colors close to 
0.4 and could therefore be more evolved pre-main sequence stars.
The group of five, possibly six sources if we include J23134708+6127090, which lie to
 the northeast of NGC\,7538\,S all have  [5.8]-[8.0] colors  $>$ 0.7, and are clearly 
young pre-main sequence stars.  The brightest 
star in the group is IRS\,11, which does not have much foreground extinction,
 [3.6]-[4.5] $\sim$ 0, but  IRS\,11 has  a strong 8 $\mu$m excess,  and is clearly a young star.
 IRS\,11\,S, the H$_2$O maser,  appears to be much more deeply embedded than IRS\,11, and it has a similar [5.8]-[8.0] color as IRS\,11.  Just north west of 
 this group is J23134459+6127148, which is surrounded by a prominent
 nebulosity at 5.8 and 8 $\mu$m and  is almost certainly a young object, but it may be associated with IRS\,1 rather than with the NGC\,7538\,S cloud core. 
 With NGC7538\,S we therefore see a young cluster with at least six sources. Our deep 1 mm CARMA images show at least two additional sub-mm sources, which must be in an even
 earlier evolutionary stage, or more deeply embedded, since they have no mid-IR counterpart.
 
% FIGURE 5
\begin{figure}[t]
\centering
  \includegraphics[width=35mm, angle=-90, origin=c]{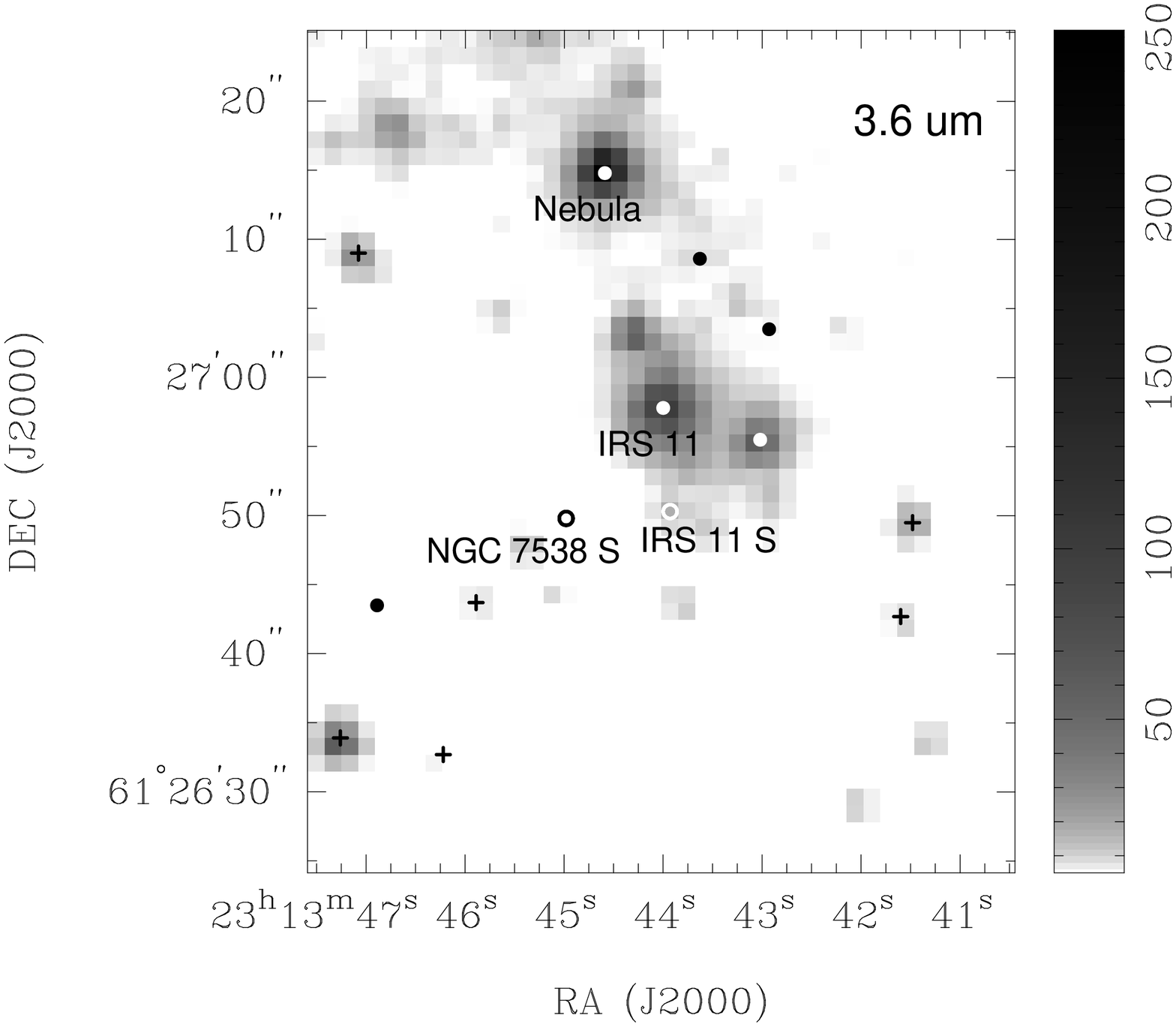}
  %\hspace*{-5mm}
  \includegraphics[width=35mm, angle=-90, origin=c]{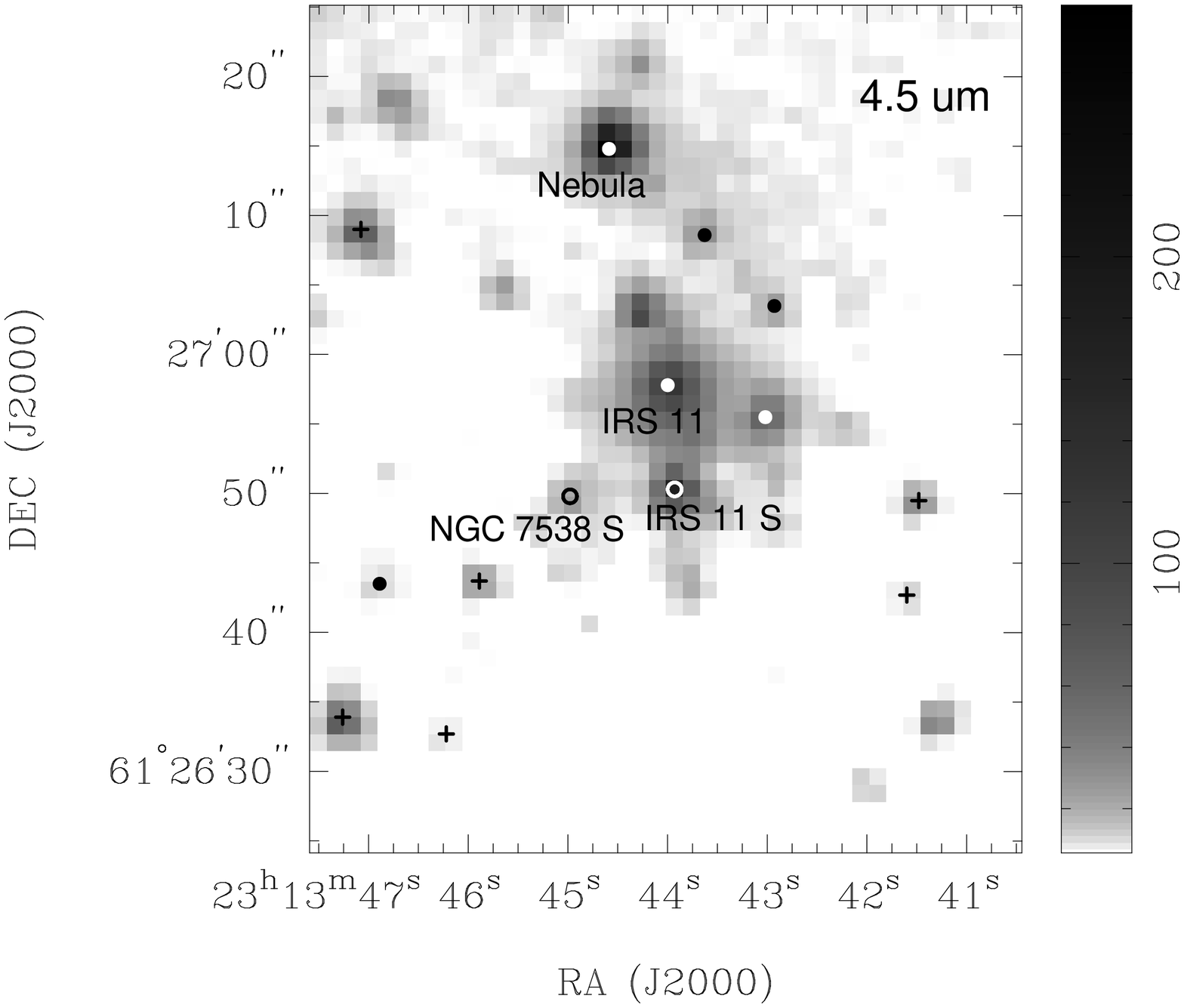}
  \\
  \includegraphics[width=35mm, angle=-90, origin=c]{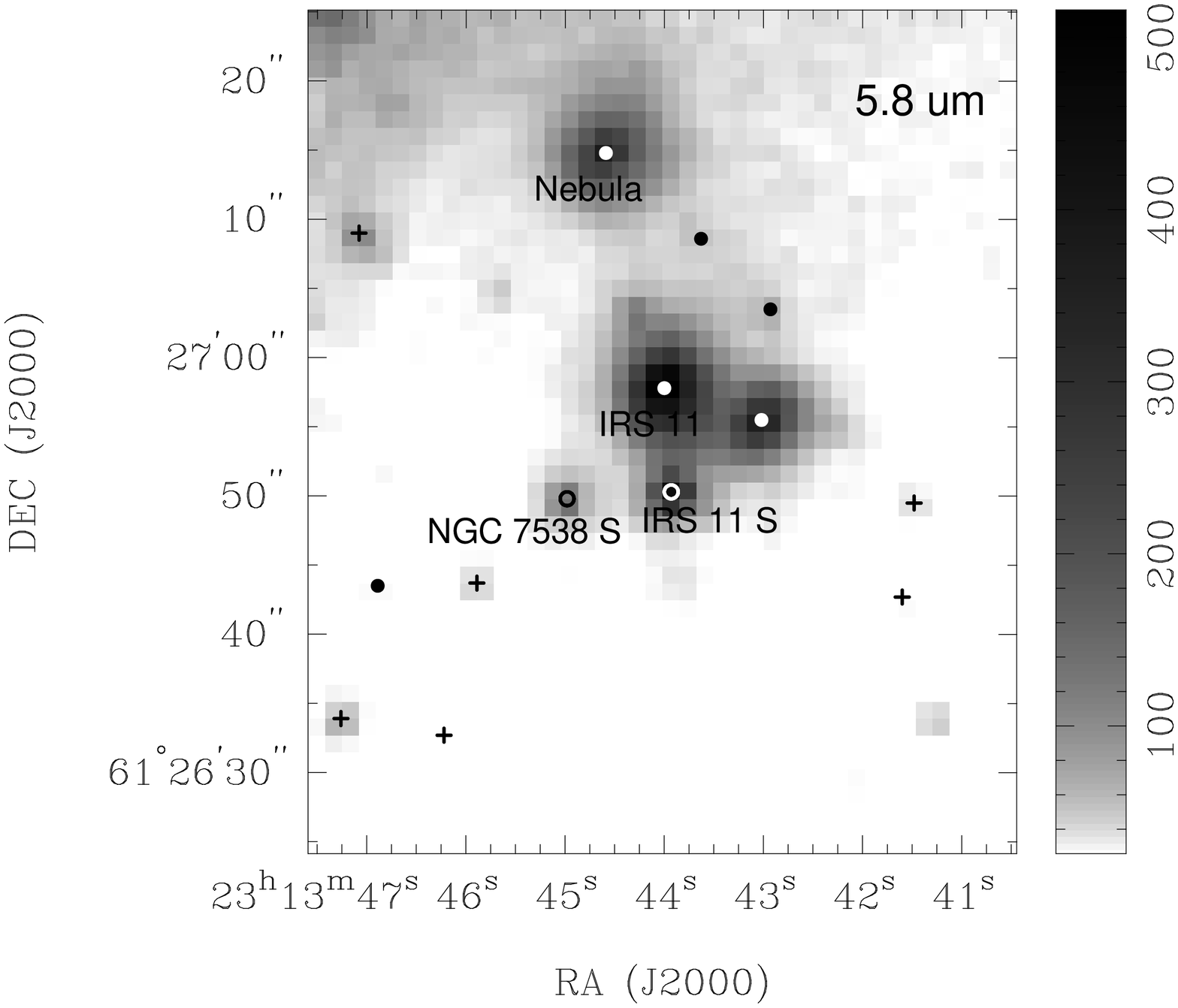}
  %\hspace*{-5mm}
  \includegraphics[width=35mm, angle=-90, origin=c]{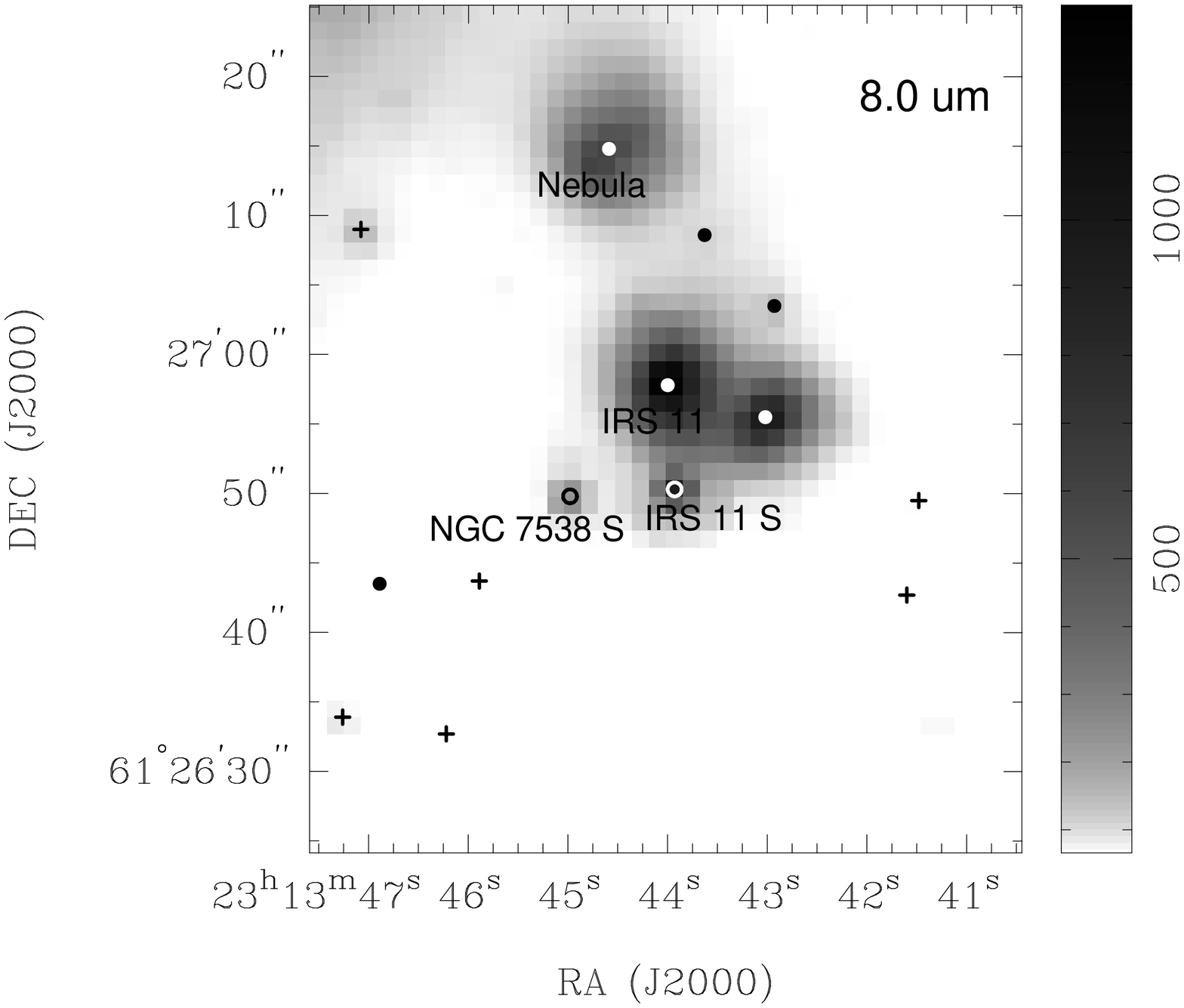}
\caption{
The detailed image of NGC\,7538\,S region in the 3.6, 4.5, 5.8 and 8.0\,\micron.
The identified point-like sources are indicated with crosses and circles. The
isolated sources labeled by crosses have relatively low color index and are
likely to be field star or evolved class III stars. The sources plotted as open
circles have  large color indeces, and are all 
young pre-main sequence objects. NGC 7538\,S and IRS\,11\,S, both of which 
were detected  in our CARMA images, are very deeply embedded ([3.6]-[4.5] $ >$
2).  NGC\,7538\,S  was not even detected at 3.6 $\mu$m.
}
\label{fig-irac}
\end{figure}

% FIGURE 6
\begin{figure}[t]
\centering
  \includegraphics[width=85mm, angle=0, origin=c]{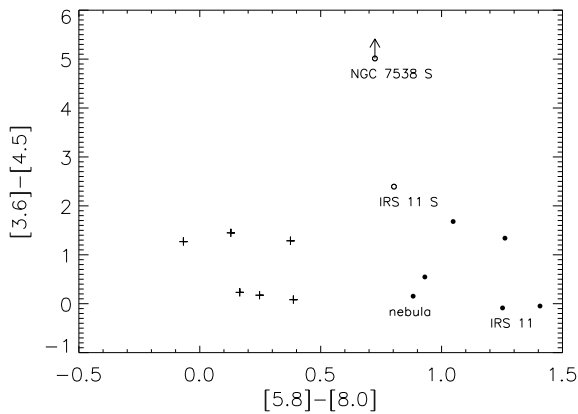}
\caption{The color-color diagram of [3.6]-[4.5] vs. [5.8]-[8.0] for
the 8\,\micron ~sources in NGC\,7538\,S. The indications of the crosses and circles
in this figuare are the same as Figure 5.}
\label{fig-c-diag}
\end{figure}

% DISCUSSION - MM
\subsection{Millimeter wavelength dust emission}
\label{mm-results}

% FIGURE 7
\begin{figure}[t]
\centering
\includegraphics[width=55mm, angle=-90]{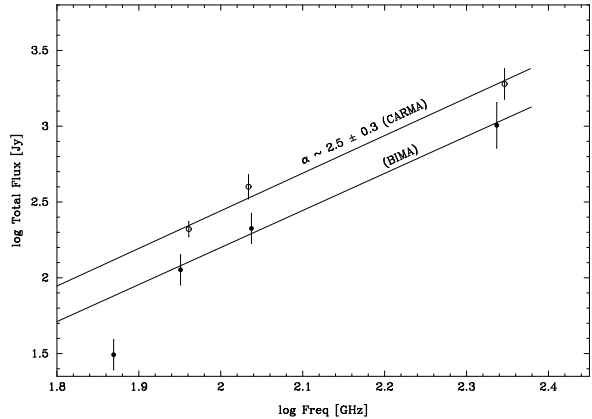}
\caption{
Integrated flux densities of NGC\,7538\,S  as a function of frequency.
These have been corrected for coherence loss (see text) and corrected
for free-free emission. The solid lines show the spectral index, $\alpha$,
determined from the BIMA and CARMA data.  The spectral index appears to
steepen at longer wavelengths. See text.
}
\label{fig-S-si}
\end{figure}

%\begin{figure*}[t]
%\centering
%\includegraphics[width=175mm, angle=0]{f6.ps}
%\caption{Spectral index image of NGC7538S.
%The spectral index values are shown at the right. }
%\label{fig-si-map}
%\end{figure*}

%The mm-SED of IRS1 \& IRS\,2 from our BIMA data confirm that our data
%are well calibrated and hence we can trust the deduced flux densities
%for NGC\,7538\,S.  
The VLA observations (Section \ref{VLA-results})
show that the free-free emission in the 3 mm band is predicted
to be $\sim$ 8.5 mJy and therefore the emission at mm-wavelengths is
completely dominated by dust emission. The free-free corrected flux
densities obtained with BIMA and CARMA for the elliptical core are plotted in Figure \ref{fig-S-si}
as a function of frequency. A least squares fit to the BIMA 89, 109 and 217
GHz data points give a spectral index, $\alpha \sim$ 2.45, which is surprisingly low. At 89 GHz,
however,  the elliptical core appears more centrally condensed and smaller (Table 2,
%\ref{tbl-2}, 
Figure \ref{fig-mmcont}), i.e. the observations are not
sensitive enough to recover  the cold outer parts of the core. Since the
89 GHz observations  do not sample the same volume of dust, it should
therefore be excluded from the fit.  Using only the flux densities
from the best images, at 109 and 217 GHz,  gives $\alpha$ =  2.3 $\pm$
0.2. This corresponds to an apparent dust emissivity, $\beta \sim$ 0.4,
if we assume that the dust is isothermal with a dust temperature of 40 K.
This is much lower than the dust emissivity derived by \citet{Sandell04}
from multi frequency single dish observations and appears suspiciously low.
%The single dish data,
%however, are likely to include emission from the surrounding cloud
%and the hot outflow. 
%Furthermore the single dish observations are all
%broadband bolometer observations which are likely to include significant
%line emission, while the BIMA continuum images exclude any line emission
%and are therefore more accurate and reliable.

%G's email 26aug2011
We therefore also examined the CARMA data presented in \citet{Corder08} as well as the new CARMA
data discussed in this paper. Corder finds a bigger size for the elliptical core, $\sim$ 12\arcsec\ $\times$ 7\arcsec, and we therefore cannot directly
compare his results to our BIMA results. Instead we compare the quoted total fluxes for the NGC\,7538\,S cloud core, 
which is 0.218 $\pm$ 0.007 Jy at 91.4GHz and 0.407 $\pm$ 0.016 Jy at 108.1GHz.  For the deep BCDE-array CARMA image at 222.2 GHz we find a total
flux for the core of 1.9 $\pm$ 0.2 Jy. A least squares fit to all three data points give $\alpha$ =  2.5 $\pm$ 0.4, which is very similar to
what we found from the BIMA data.
% should 108 be 111 here ?
If we only use the 108 and 222 GHz we get $\alpha$ =   2.2 $\pm$0.4. 
Both the CARMA and the BIMA data  (Fig. 7) show
a steepening of the spectral index at long wavelengths. If we only use the data points at 91 and 108 GHz, we get $\alpha$  = 3.8 $\pm$ 0.5, which is consistent with thermal emission from dust.
What these data therefore suggest is that the elliptical core may be so extreme, that the dust is already becoming optically thick at  1 mm. It is possible that what
we observe is partly an instrumental effect, i.e., we filter out more emission at the highest frequencies. However,  the same trend is seen
in both our BIMA and our high fidelity CARMA data, which at 222.2 GHz have extremely good uv-coverage. This suggests that the trend is real and not just an artifact 
from different response  in spatial sensitivity for an aperture synthesis array at different frequencies.

One would in fact
expect the surrounding cloud core to have a steeper spectral index than
the disk, because most studies find that the emissivity index $\beta$
$\sim$ 1.5 - 2 \citep{Wright92,Masi95,Goldsmith97}, i.e. corresponding
to $\alpha$ $\sim$ 4.  The most likely explanation to the observed low
spectral index of the surrounding cloud core is that the BIMA array
filters out much more of extended emission at 217 GHz than at 109 GHz.
The single dish observations by \citet{Sandell04} shows that the cloud
core surrounding NGC\,7538\,S has a radius of $\sim$  30\arcsec, while
our interferometer observations find a core radius of less than half
of this value, suggesting that most of the cloud core is resolved out
in the BIMA observations. Both the 217 and 109 GHz observations 
recover spatial scales of \ltsim 20\arcsec; therefore the results for
the disk surrounding NGC\,7538\,S should not be affected.

%f9.cps is the high resolution image extended to the west to include the IRAC source.
%\includegraphics[width=45mm, angle=-90]{f9.cps}

%FIGURE 8
\begin{figure}[t]
\centering
\includegraphics[width=45mm, angle=-90]{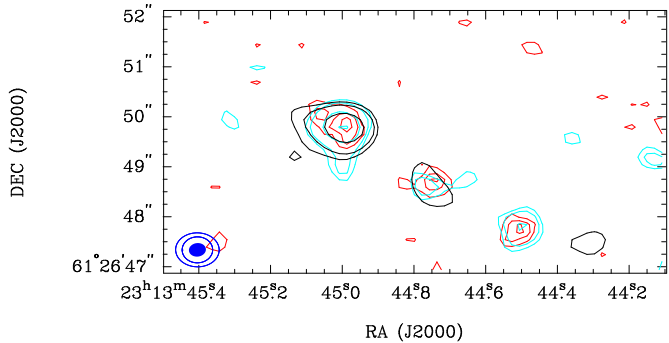}
\caption{
High spatial resolution images of NGC\,7538\,S obtained 
with CARMA at 88 (black), 111 (blue) and 222 (red) GHz. 
The synthesized beam FWHM shown in the lower 
left corner are 0.87\arcsec$\times$0.68\arcsec,  
0.61\arcsec$\times$0.52\arcsec, 
and 0.34\arcsec$\times$0.27\arcsec  at  88,  111 and 222 GHz respectively.
Contour intervals: 2.56 mJy/beam at 222 GHz,  2, 3, 6, 9 mJy/beam at 111 and 88 GHz.
}
\label{carmahr-image}
\end{figure}

Although variations in the observed spectral index or dust emissivity
are usually interpreted as changes in dust composition, such changes
can also be caused by temperature and density gradients if the dust
is optically thick.  \citet{Beckwith91} found that most T Tauri stars
have spectral indeces between $\sim$ 2 - 3, which they showed to result
from T Tauri stars having optically thick inner disks at millimeter
wavelengths. \citet{Dutrey96} did a more careful study using the IRAM
interferometer at 2.7 mm and confirmed that all the stars in her sample
had optically thick inner disks.  It is now also well established that
disks around more evolved T Tauri and Herbig Ae stars have low dust
emissivity, but  for these stars the drop in emissivity is due to grain
growth, not optically thick dust  \citep{Natta04}.

% FIGURE 9
\begin{figure}[t]
\centering
\includegraphics[width=85mm, angle=0]{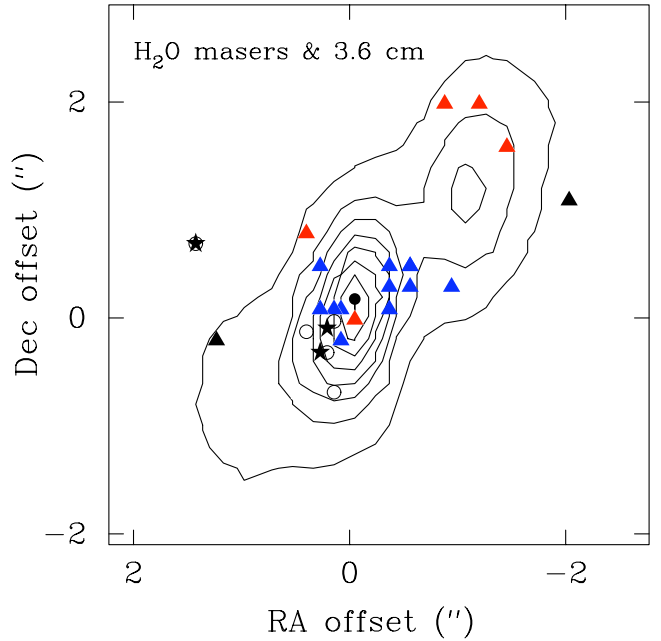}
\caption{
Overlay of H$_2$O maser positions on the VLA 3.6~cm continuum image. The contours for the continuum 
image are linear with seven  contours between 0.10 mJy~beam$^{-1}$ and the peak flux, 2.47 mJy
~beam$^{-1}$. The  H$_2$O maser spots are marked with triangles, where the red triangles show red
 shifted maser spots, the blue ones show  blue shifted masers, and the black ones are near the sy
stemic velocity of the cloud. The filled circle shows the position of the IRAC mid-infrared sourc
e. The filled star symbols are RCP (right circularly polarized) OH 1665 maser spots, while the open circles are LCP OH 1665 maser spots from \citet{Argon00}, which are measured to an accuracy of
 0\farcs01. The cross marks the position of the CH$_3$OH  class II maser from \citet{Pestalo06},
who quote an astrometric accuracy of  0\farcs5.
Position offsets are w.r.t. NGC\,7538\,$S_a$ 
}
\label{valmaser-image}
\end{figure}

For high-mass protostellar objects, single dish surveys show a much
larger scatter in the observed dust emissivity. \citet{Williams04} did
a SCUBA  850 and 450 $\mu$m survey of 68 protostellar objects and found
a mean spectral index, $\alpha$ = 2.6 $\pm$ 0.4, corresponding to a dust
emissivity, $\beta$ = 0.9 $\pm$ 0.4, suggesting that the dust emissivity
in high-mass protostellar objects has largely the same characteristics as
dust disks surrounding T Tauri and Herbig Ae/Be stars.  On the other hand,
\citet{Hill06}, who  did a similar survey using SCUBA observations at
850 and 450 $\mu$m and SIMBA observations at 1.3 mm of 212 cold massive
cloud cores,  argue that their results are in agreement with a typical
$\beta$-value of 2, although their results show  considerable scatter in
the derived $\beta$ indeces to both higher and lower $\beta$-values.

However, all published millimeter array observations of high-mass
protostellar objects find consistently low $\beta$ values. For the
accretion disk in IRAS $20126+4104$, which is barely resolved in
mm-continuum, \citet{Cesaroni05} find a spectral index of 2.7 from
a least squares fit to all published Plateau de Bure  Interferometer
(PdBI) data at frequencies 86.6, 90.6, 96.5,  217.6  and 239.6 GHz,
the Owens Valley Radio Observatory (OVRO) millimeter wave array data at
113.2 and  228.1 GHz, and a 7 mm  (42.8 GHz) data point from the VLA.
\citet{Beuther04a} were able to resolve the high mass protostellar core
mm1 in IRAS $20293+3952$ with PdBI at  98.0 and 244.9 GHz, and found a
spectral index of $\sim$ 3.4 at the outer edge of the core decreasing
to values as low as $\sim$ 2.1 at the very center of the core.
SMA observations of the massive protostellar core IRAS $18089-1732$
with the  SMA at 217 and 354 GHz \citep{Beuther04b} give similar
results from analysis of fluxes in the u-v plane; $\beta$ $\sim$  1.2
for large spatial scales, 5\arcsec\ - 10\arcsec, and decreasing to $\sim$
0.5 for small spatial scales, 2\ptsec7 - 3\ptsec6. Therefore, in the few
cases where high-mass protostellar objects have been resolved,  they all
indicate that the dust emissivity drops towards the center of the core,
which is believed to be dominated by emission from an accretion disk.

%Even though there is at least one  early B star (R Mon, spectral class B0)
%with a disk, which shows strong evidence for grain growth \citep{Fuente06},
The high-mass protostellar objects discussed here are very young. It
is therefore  unlikely that grain growth would have had time to
occur. A more plausible explanation for our low spectral index between
111 and 222 GHz is that the disks are optically
thick at mm-wavelengths. Our results on NGC\,7538\,S$_A$ strongly favor high
optical depth, rather than a population of large grains in the disk. It
is clear from the BIMA images (Figure \ref{fig-mmcont}) that
S$_A$ starts to dominate the emission of the elliptical core at the longest wavelengths, suggesting that one can now see the center of  the accretion disk, which is expected to
be hotter and denser than the surrounding infalling envelope. The center of the
disk therefore dominates the emission.

If the disk is optically thick at 111 and 222 GHz, as our observations
suggest, one would really need to model the dust emission in order to derive a
disk mass. However, we can get an estimate of the disk mass by assuming that the
dust emission is optically thin and roughly isothermal at the longest wavelength
for which we have a good high resolution image, i.e. at 89.3 GHz. The total mass
of gas and dust, M,  can then be expressed as  ${\rm M = S_\nu D^2/(\kappa_\nu
B_\nu(T_d))}$, where ${\rm B_\nu(T_d)}$ is the Planck function, $ \rm T_d$ is
the dust temperature, D is the distance, and S$_\nu$ is the integrated flux
density at the frequency $\nu$. Because of the uncertainty in dust emissivity,
we do not want to use  the \citet{Hildebrand83} dust opacity and extrapolate it
from  250 $\mu$m. Instead we have chosen to use  the dust mass opacity, $\kappa$
as tabulated by \citet{Ossenkopf94} at 1.3 mm. Their results suggest a dust mass
opacity  $\kappa_{\rm 1.3 mm}$ = 0.01 ${ \rm cm^2 g^{-1}}$ for gas densities of
10$^8${ \rm cm$^{-3}$} and grains with thick ice mantles,
assuming the ratio of gas to dust mass is 100. These are the type of
dust grains and densities that one would expect to have in the NGC\,7538\,S
accretion disk. At  frequency $\nu$, the dust mass opacity is therefore
$\kappa_\nu = 0.01 \times (230/\nu)^{-\beta}$  ${ \rm cm^2 g^{-1}}$. At  91.4
GHz we now obtain a dust mass opacity of  $\sim$ 0.0025 ${ \rm cm^2 g^{-1}}$  if
 we assume $\beta$ = 1.5, which is probably appropriate for this case.  If we further assume a
dust temperature of 35 K, see below,  a distance 2.65 kpc,  and correct the observed flux density at 91.4 GHz for
free-free emission ($\sim$ 8.4 mJy) we obtain a disk mass  $\sim$ 60 \Msun.
This is about two times smaller than what was reported in \citet{Sandell03}, 
but in reasonable agreement with the dynamical mass derived from H$^{13}$CN
\jtra{1}{0},  20 \Msun\  \citep{Sandell05}. Since the size of the rotationally 
supported part of the disk is
smaller than what we measure in continuum, we would expect 
the mass derived from continuum to be higher.

The assumption of a dust temperature of 35 K appears justified, see e.g.  \citet{Sandell04} and \citet{Sandell10}.
\citet{Zheng01} find ammonia to be optically thick in the direction of NGC\,7538\,S with a temperature of 25 K. The temperature of the disk is
likely to be less than 50 K, because then we would exceed the observed far infrared luminosity \citep{Thronson79}.  The dust temperature must be higher than 25 K, the temperature of the
surrounding cloud core, since  \citet{Sandell10}  see some self-absorption in virtually
every molecule seen in  the disk.  We can get another upper limit by looking at the
analysis of the CH$_3$CN transitions by \citet{Sandell10}, where they derive a rotation temperature = 52 $\pm$ 10 K. Since the  low K-transitions are  optically thick \citep{Sandell10}, this leads to an
overestimate of the rotational temperature. Contribution from the hot
outflow will also raise the temperature. Therefore a dust temperature of 50 K  appears to be a reasonable upper limit.
If we were to increase the dust temperature  to 50 K, this would change our mass estimate by less than a factor of two.

We can also do an order of magnitude  estimate of the disk mass if we assume 
that the inner disk is optically thick at 1.4 mm using a geometrically thin 
disk model, which has been successfully used to analyze millimeter continuum 
emission from T Tauri stars \citep{Beckwith90, Dutrey96}.  For an optically 
thin disk the optical depth, $\tau_\nu =  \kappa_\nu \times \Sigma(r)/cos(\theta)$, 
where $\kappa_\nu$ is dust mass opacity, $\Sigma(r)$ is the surface density 
as a function of radius, r, and $\theta$ is the inclination angle of the disk 
($\theta$ = 90\degr\ for an edge on disk). The surface density is assumed to 
be a power law of radius, i.e. $\Sigma(r) = \Sigma_o \times (\displaystyle \frac{r}{R_o})^{-p}$. 
If we define the disk radius, R$_{\rm D}$, as the radius where the optical 
depth, $\tau_\nu$ = 1, we can express the disk mass M$_{\rm D}$ as 

%%JHZ: I deleted \Sigma_0 in  equation (1)
\begin{equation}
      M_{\rm D} =  \frac{2\pi}{2 - p} \frac{cos(\theta)}{\kappa_\nu} R_{\rm D}^2
\end{equation}

\noindent
which, if we adopt a dust mass opacity of 0.009 ${ \rm cm^2 g^{-1}}$ at 217.3 GHz gives:

\begin{equation}
         M_{\rm D} =  \frac{78.6}{2 - p} cos(\theta) (\frac{R_{\rm D}}{1000 \rm{AU}})^2  ~~~~  [M_\odot]
\end{equation}

For an optically thick radius of 0\ptsec5 (1400 AU), roughly the size
measured with CARMA,  a relatively flat mass 
distribution, i.e, p = 1, and a disk inclination  of 70\degr\ , we 
derive a  mass of  $\sim$ 50  \Msun\ for the optically 
thick disk component, which is similar to what we estimate assuming the
dust is optically thin at 91.4 GHz. The radius of the optically thick disk is
probably somewhat smaller, so this is in fact an upper limit. In order to be able to 
determine the radius of the optically thick inner disk, 
it would be necessary to have observations with a resolution of 0\farcs1 - 0\farcs2  at at least three frequencies, one of which would need to be below 100 GHz. Such observations are now 
completely feasible.

We can estimate the dust opacity towards  S$_A$ from our high resolution CARMA data. 
 We need to subtract the emission from the thermal jet, which from the fit to our VLA data is predicted to be:
        8.3, 8.6, and 9.5 mJy  at 87.9, 111.1, and 222.2 GHz respectively.
        Fitting the three data points at the above frequencies on S$_A$
after subtracting the contribution from the free-free emission
(here we add the flux for S$_{A1}$ and S$_{A2}$ at 222.2 GHz ,and  the flux is  assumed to be 47 mJy),
we obtained a spectral index $\alpha=2.1\pm0.3$, assuming that all data points have a 10\% error.
This is a  strong indication that the dust emission towards S$_A$, our high mass protostar, must be optically thick.   
Because the dust emission from the NGC\,7538S$_A$ disk is rather 
strong, it should now be possible to detect the dust emission 
with the eVLA both at 22 and 43 GHz, which would allow us 
to get a better estimate of the dust emissivity index.

\subsection{Protostars NGC\,7538\,S$_{\rm A}$, S$_{\rm B}$, and S$_{\rm C}$}
In addition to NGC\,7538\,S$_{\rm A}$,
% that is associated with the ionized outflow detected with the VLA, 
 the high-resolution CARMA image (Figures \ref{224-BCDE-S-sdicm-regrid.pdf}
and \ref{carmahr-image})  reveals two compact components (S$_{\rm B}$ and S$_{\rm C}$) 
southwest  of S$_{\rm A}$ along the ridge of the accretion disk. No radio continuum emission 
from either of these two components has been detected with the VLA, suggesting that they 
are in an early protostellar phase. They may have been produced from fragmentation due to 
an instability of the material in the accretion disk of  NGC\,7538\,S.    
They are the youngest members in the young stellar object (YSO) group of 
NGC 7538S including IRS sources detected with Spitzer.
%G's version:
When the elliptical core  started collapsing, S$_A$ was probably the first protostar to be 
surrounded by a rotating disk which already at that point was very dense and optically thick. 
We know that there is a velocity gradient over the whole $8''$ elliptical core. Whether S$_B$ and S$_C$ fragmented from the massive, unstable disk around S$_A$ or whether the core fragmented into three protostars is unclear. It would be important to know but it is beyond the scope of our current paper.
We see a very dense rotating core in which we find three extremely young protostars. S$_A$ has the luminosity and mass to form a star which may evolve into an early B or late O star. We do not know the luminosity of the other two, but judged on the $\lambda$mm flux (and they have no free-free), they are similar or slightly lower  mass, so they probably will form mid-- to late B-stars.
The  Bonner-Ebert mass estimated from the observed size and flux density at 225 GHz for dust temperatures 25 to 50 K are 0.01 to 0.1\Msun\ .  The mass of the compact sources exceeds the Bonnor-Ebert mass, suggesting that they are local accretion centers \citep{Bonnor56, Ebert57}.
Competitive accretion will make S$_A$ grow faster, so it will almost certainly be the most massive star. The accretion flow, within the errors, is centered  on what we now know to be $S_A$ \citep{Sandell10}. As we have shown in this paper this is also the most luminous and massive  protostar in the elliptical core. It is therefore reasonable to
assume that accretion will continue until $S_A$ forms an HII region, which then will halt the accretion onto the star.  However, when this happens, and after the HII region expands, it will also halt the accretion onto the nearby protostars, so $S_A$ will remain the most massive star.  

\section{Conclusion}
We used BIMA and CARMA observations of NGC 7538S at millimeter
wavelengths, VLA observations at centimeter wavelengths, and Spizter in the mid IR bands, to study the high-mass star formation in this YSO group. 
The observations at millimeter wavelengths show that the continuum emission from NGC
7538\,S is dominated by dust emission from an  8\arcsec$\times$3\arcsec source, elongated  
 northeast to southwest, suggesting an accretion disk. 
 The observed spectral index of $\alpha\sim2.3$
implies that the dust in the disk either has a low dust emissivity or is optically thick. 

The elliptical core breaks up into three $\lambda$mm sources.  
The brightest one, NGC\,7538\,S$_A$, is extended and is possibly a binary. 
This $\lambda$mm source coincides with an OH 1665 MHz maser, a Class II methanol maser, a cluster of H$_2$O maser spots, a faint free-free
VLA jet, and a heavily obscured mid-IR source ({\it Spitzer} IRAC and IRS data). 
It drives a very compact bipolar outflow and appears to be
surrounded by a rotating disk with a radius of $\sim$2\arcsec\ \citep{Sandell10}. 
The other two $\lambda$3 mm sources along the central ridge of
the clump both appear point-like, although one of them may drive another compact outflow \citep{Corder08}.

NGC 7538S$_{\rm A}$ appears to be the most
powerful source in this group and produces an ionized bipolar outflow, detected by the VLA,  in a direction perpendicular to the major axis of the disk.
At millimeter wavelengths, with sub-arcsec resulution, 
CARMA has resolved this source (S$_{\rm A}$) 
into two sub-components distributed along the major axis of the accretion disk.
Located southwest of S$_{\rm A}$ along the major axis
of the accretion disk, two compact millimeter sources S$_{\rm B}$
and S$_{\rm C}$ are also detected at 222 GHz with CARMA, 
suggesting that they may be the results from the fragmentation of the 
accretion material in NGC 7538S. Along with the highly reddened objects
detected with Spitzer at  mid IR bands, the protostars detected at millimeter wavelengths
suggest that NGC 7538S consists of a group of very young stellar objects
in the NGC 7538 complex.
\acknowledgements 
The BIMA array was operated by the Universities of California (Berkeley),
Illinois, and Maryland with support from the National Science
Foundation.
Support for CARMA construction was derived from the 
states of California, Illinois, and Maryland, the Gordon 
and Betty Moore Foundation, the Kenneth T. and Eileen 
L. Norris Foundation, the Associates of the California 
Institute of Technology, and the National Science Foun- 
dation. Ongoing CARMA development and operations 
are supported by the National Science Foundation under 
a cooperative agreement, and by the CARMA partner 
universities.
%The Very Large Array (VLA) is operated by the National Radio Astronomy
%Observatory (NRAO). The NRAO is a facility of the National Science
%Foundation operated under cooperative agreement by Associated
%Universities, Inc.

\appendix
\section{Atmospheric decorrelation}

The resolution and dynamic range of millimeter wavelength aperture
synthesis images are limited by atmospheric turbulence.  Atmospheric phase
noise reduces the observed amplitudes by decorrelating the signals
received at each antenna.  In an analysis of data obtained with
the Hat Creek interferometer at 86 GHz we found a baseline dependent
phase structure function with a power law index, $\beta$ between 0.6
and 1.7.  A slope close to the Kolmogorov 2D value 0.67 is obtained for
longer baselines and stable weather conditions; the slope approaches the
Kolmogorov 3D value 1.67 on short baselines and in turbulent weather.
The rms pathlength on a 1 km baseline $\sim$ 1 mm, with a variation
by a factor 4 over a few days even during good weather conditions \citet{Wright96}.

We estimated the atmospheric decorrelation for the NGC\,7538  observations
by analysing the phase calibrator data.  The phase calibrator was
observed at $\sim$ 25 minute intervals and samples the same atmospheric
fluctuations as the observations of NGC7538. Although the short calibrator
observations do not have sufficient signal-to-noise to determine
the radio seeing for each observation of NGC7538, we can estimate the
seeing over several hours from the observed calibrator phase scatter.
Atmospheric phase noise increases the measured FWHM of the calibrator image.
We calculated amplitude corrections for atmospheric decorrelation by
fitting a circular Gaussian model to images of the phase calibrator.
% made using only the calibrator phase. 
 
The BIMA  data for  NGC\,7538 were corrected by
applying an amplitude correction for atmospheric coherence, by specifying
the FWHM in arcsec, of a circular Gaussian fit to the phase calibrator(s)
for the target source.  The derived scaling factors from decorrelation
loss were 1.11 at 74 GHz, 1.08 at 89 GHz, 1.50 at 109 GHz, and  1.18
at 217 GHz. The flux densities in Table 1 have been corrected for
these decorrelation losses. 
%The corrected fluxes for IRS\,1 and IRS\,2
%are plotted in {\it Figure 3 which has not inluded?} together with 
%data from \citet{Akabane92} and
%\citet{Tak00} and agree very well internally as well as with published values.

{}

\clearpage
\begin{deluxetable}{lcclcl}
\tabletypesize{\scriptsize}
\tablecolumns{6}
\tablenum{1}
\tablewidth{0pt} 
\tablecaption{BIMA observing log. \label{tbl-1}} 
\tablehead{
\colhead{Array Configuration}  & \colhead{rest frequency}  & \colhead{Bandwidth} & \colhead{Synthesized beam} & \colhead{rms} & \colhead{Observing dates}\\ 
\colhead{} & \colhead{[GHz]} & \colhead{[MHz]} &\colhead{\arcsec{} $\times$ \arcsec{} pa = \degr{} }  & \colhead{[mJy~beam$^{-1}$]} & \colhead{}
%\cline{1-6} 
}
\startdata
\sidehead{Frequency setting: DCN \jtra10}
\tableline \\[-1mm]
BCD & 73.9 & 400 & 9.1 $\times$ 7.2 ~~78.9  & 3.7 & 22 Jul \& 30 Oct 03, 07 Jan 04 \\[2mm]
\sidehead{Frequency setting: HCO$^+$ \jtra10, H$^{13}$CN  \jtra10, NH$_2$D  1$_{11}  \to  1_{01}$, SO  \jtra{2,2}{1,1}{}} 
\tableline\\[-1mm]
BCD & 89.3 & 75  & 4.1 $\times$ 3.9 ~~~8.8  & 3.0 & 27 \& 31 Jan ,\& 03 May 02\\
         &           &                &                                                            &       & 09 Aug, 04 Nov, 11, 12, \& 31 Dec 03\\ 
\sidehead{Frequency setting: Continuum}
\tableline\\[-1mm]
BC & 109.0 & 800 & 3.5 $\times$ 3.2 ~~~89.5 & 0.54 & 27 Mar \& 27 Apr 04 \\
\sidehead{Frequency setting: DCN \jtra32, $^{13}$CO \jtra10, CH$_3$CN \jtra{12}{11}}
\tableline\\[-1mm]
BC & 217.3 & 150 & 2.9 $\times$ 2.4 ~~~78.4 &  3.0 & 13 Oct 02, 19 Apr, 1 May 03 \\[2mm]
  &   &   &   &  & 10, \& 25 Oct 03, 3, \& 5 Jan 04
\enddata
\end{deluxetable}
 
\clearpage

\begin{deluxetable}{llllccc}
\tabletypesize{\scriptsize}
\tablecolumns{6}
\tablenum{2}
\tablewidth{0pt} 
\tablecaption{Continuum positions and flux densities for sources in  the NGC\,7538\,S core from VLA, BIMA and CARMA observations. \label{tbl-2}}
\tablehead{
\colhead{Source} &
\colhead{Frequency} & 
\colhead{$\alpha$(J2000.0)} & 
\colhead{$\delta$(J2000.0)} & 
\colhead{$\theta_a\times\,\theta_b$\tablenotemark{a}} & 
\colhead{p.a.}  &
\colhead{S$_{int}$} \\
\colhead{} & 
\colhead{[GHz]} & 
\colhead{[$^h\,^m\,^s$]} & 
\colhead{[$^\circ$\,\arcmin\,\arcsec ]} &  
\colhead{[\arcsec$\times$\arcsec]} & 
\colhead{[$^\circ$]}  & 
\colhead{[mJy]} }
\startdata
\multicolumn{7}{c}{} \\
\multicolumn{7}{c}{VLA Observations} \\
\multicolumn{7}{c}{} \\
NGC\,7538\,S\,\,a&4.86 & 23 13 44.992 & $+$61 26 49.6 & 0.9 $\times$ $<$0.3        & $+$131&\phantom{0}3.60 $\pm$ 0.18\\
\phantom{NGC\,7538\,S\,\,}b& & 23 13 44.817 & $+$61 26 50.9 & 0.4 $\times$ $<$0.5        & $+$119&\phantom{0}1.76 $\pm$ 0.17\\
% tvstat gives 4.50 ; sum of Gaussians  5.36
\phantom{NGC\,7538\,S\,\,}a&8.46& 23 13 44.971 & $+$61 26 49.7 & 0.6 $\times$ $<$0.1        & $+$137&\phantom{0}3.25 $\pm$ 0.09\\
\phantom{NGC\,7538\,S\,\,}b& & 23 13 44.822 & $+$61 26 50.8 & 0.9 $\times$ \phantom{<}0.4& $+$136&\phantom{0}1.69 $\pm$ 0.11\\
\phantom{NGC\,7538\,S\,\,}c&& 23 13 45.066 & $+$61 26 49.1 & 1.0 $\times$ $<$0.1        & $+$117&\phantom{0}0.93 $\pm$ 0.12\\
 % tvstat 5.756  ;sum of Gaussians  6.29
&14.9     & 23 13 44.953 & $+$61 26 49.8 & 0.45$\times$ 0.24          & $-$36 &\phantom{0}4.4\phantom{0}                                                                             $\pm$ 0.4\tablenotemark{a} \\
\phantom{NGC\,7538\,S\,\,}a&22.49 & 23 13 44.978 & $+$61 26 49.7 & 0.8 $\times$ 0.1           & $-$37 &\phantom{0}4.77 $\pm$ 0.06\\
\phantom{NGC\,7538\,S\,\,}b& & 23 13 44.797 & $+$61 26 51.2 & 1.2 $\times$ 0.4           & $-$29 &\phantom{0}1.98 $\pm$ 0.08\\
 % sum of the two components          6.75
\multicolumn{7}{c}{} \\
\multicolumn{7}{c}{BIMA Observations} \\
\multicolumn{7}{c}{} \\
{NGC\,7538\,S\,\,}&74.0 & 23 13 44.836 & $+$61 26 50.5 & point-source\tablenotemark{b}&\nodata&\phantom{0}39.3$\pm$4.0\phantom{0}\\
% multiply by 1.11 * 35.4
&89.3     & 23 13 44.826 & $+$61 26 49.1 & 7.4 $\times$ 3.4         & $+$70 &\phantom{0.}121$\pm$12\phantom{.0} \\
%multiply by 1.08 *80.7 =
%maybe use this fit  1.08 * 112.3
&109.0    & 23 13 44.916 & $+$61 26 48.8 & 8.2 $\times$ 3.9         & $+$48 &\phantom{0.}220$\pm$20\phantom{.0} \\
%multiply by (1.5 *  146.9 = 220.4)
&217.2    &23 13 44.909&$+$61 26 48.8&7.8$\times$2.8&$+$54&\phantom{0.}1020$\pm$100\phantom{0.}\\
%multiply by 1.17 * 874.6 = 1023.3
%at 89  16.3
%multiply by (1.5 * 22.2
\phantom{MMMMM}South &109.0 &23 13 44.20\phantom{0}&$+$61 26 42.2 & 5.1$\times$2.8&\nodata&22$\pm$4\phantom{0}\\
%multiply by 1.17 * 95.5
&217.2    &23 13 44.28\phantom{0}&$+$61 26 42.2 & 5.5$\times$0.1 &\nodata& 96$\pm$10 \\
%\sidehead{NGC\,7538\,S West}
% at 89      1.89
%multiply by (1.5 * 30.47
\phantom{MMMMM}West &109.0&23 13 44.12\phantom{0}&$+$61 26 50.2& 6.7$\times$2.7 &$+$25$\pm$1& 30$\pm$3\phantom{0}\\
%multiply by 1.17 * 155.3
&217.2&23 13 44.18\phantom{0}&$+$61 26 50.0&4.3$\times$2.8&$+$57$\pm$3&155$\pm$20\phantom{0}\\
\multicolumn{7}{c}{} \\

\multicolumn{7}{c}{CARMA B-array Observations} \\
\multicolumn{7}{c}{} \\
\phantom{MMMMM}S$_{\rm A2}$ &222.2&  23 13 45.057&61 26 50.1 &0.49$\times$0.33 & 35 & 26.1 \\
\phantom{MMMMM}S$_{\rm A1}$ &222.2&  23 13 44.986&61 26 49.8 &0.51$\times$0.34 &-22 & 40.9 \\
\phantom{MMMMM}S$_{\rm B}$  &222.2&  23 13 44.750&61 26 48.8&\nodata&\nodata&36 \\
\phantom{MMMMM}S$_{\rm C}$  &222.2&  23 13 44.510&61 26 47.8 &\nodata&\nodata&34 \\
\phantom{MMMMM}S$_{\rm D}$  &222.2&  23 13 43.88\phantom{0} &61 26 50.2 &\nodata&\nodata& \ 5 \\
\phantom{MMMMM}S$_{\rm A}$  &111.1&  23 13 45.005&61 26 49.8 & 0.61$\times$0.58& 48.7&20.1 \\
\phantom{MMMMM}S$_{\rm B}$  &111.1&  23 13 44.750&61 26 48.8 &\nodata&\nodata&5  \\
\phantom{MMMMM}S$_{\rm C}$  &111.1&  23 13 44.510&61 26 47.8 &\nodata&\nodata&8 \\
\phantom{MMMMM}S$_{\rm D}$  &111.1&  23 13 43.878  &61 26 50.2 & 1.79$\times$0.68& 20.1 & 17.5  \\
\phantom{MMMMM}S$_{\rm A}$  & 87.9&  23 13 45.005&61 26 49.8 & 0.84$\times$0.37&85.5&13.3\\
\phantom{MMMMM}S$_{\rm B}$  & 87.9&  23 13 44.750&61 26 48.8 &\nodata&\nodata&3 \\
\phantom{MMMMM}S$_{\rm C}$  & 87.9&  23 13 44.510&61 26 47.8 &\nodata&\nodata&2\\
\phantom{MMMMM}S$_{\rm D}$   & 87.9 &  23 13 43.88\phantom{0}  &61 26 52.2 &\nodata&\nodata& \nodata  \\
\enddata
\tablenotetext{a}{At the edge of the primary beam (correction factor $\sim$ 2). Flux density underestimated because the observations are not sensitive enough to pick up faint extended emission in a 0\ptsec1 beam.}
\tablenotetext{b}{Compact or marginally resolved point-like source in 9\arcsec\ beam embedded in faint extended emission.}
\end{deluxetable}

\clearpage

\begin{deluxetable}{lllrrrrl}
\tabletypesize{\scriptsize}
\tablecolumns{8}
\tablenum{3}
\tablewidth{0pt} 
\tablecaption{Positions and flux densities of 8 $\mu$m IRAC  sources in the vicinity of NGC\,7538\,S.\label{tbl-3}}
\tablehead{
\colhead{Source} & \colhead{$\alpha$(2000.0)} & \colhead{$\delta$(2000.0)}  & \colhead{S(3.6 $\mu$m)}& \colhead{S(4.5 $\mu$m)} & \colhead{S(5.8 $\mu$m)} & \colhead{S(8 $\mu$m)}  &  \colhead{Comment} \\ 
   & \colhead{[$^h$  $^m$ $^s$]}& \colhead{[$^\circ$ \arcmin\ \arcsec ]}& \colhead{[mJy]} & \colhead{[mJy]}  & \colhead{[mJy]}  & \colhead{[mJy]}  & \colhead{}
}
\startdata
J23134726+6126339 & 23 13 47.26  & $+$61 26 33.9 & 10.19 (0.04)  & 12.61 (0.07)   & 11.68 (0.09)   & 13.6 (0.3) & field star \\
J23134708+6127090 & 23 13 47.08  & $+$61 27 09.0 & 5.07 (0.03) & 16.57 (0.17) & 20.60 (0.16) & 29.1 (0.4) &  \\
J23134689+6126435 & 23 13 46.89  & $+$61 26 43.5 & 0.59 (0.01) &  2.24 (0.02)   & 3.11 (0.05) &  3.5 (0.3)  & field star\\
J23134622+6126327 & 23 13 46.22  & $+$61 26 32.7 & 0.81 (0.01)   & 1.34 (0.02)  &  1.57 (0.04)  & 3.7 (0.3) &   \\
J23134589+6126437 & 23 13 45.89  & $+$61 26 43.7 & 1.55 (0.01)      &    4.99 (0.04)       &    7.45 (0.06)    &  7.0 (0.3) & field star \\
 J23134498+6126498 & 23 13 44.98  & $+$61 26 49.8  & $<$0.05\phantom{a(0.01)}  & 5.06 (0.03)  & 32.23  (0.15) & 62.8 (0.6) & NGC\,7538\,S\\
 J23134459+6127148 & 23 13 44.59 & $+$61 27 14.8 & 48.09 (0.32) &  55.39 (0.27) & 102.8\phantom{0} (0.6\phantom{0}) &  231.7 (1.1) & nebulous object \\
 J23134400+6126578 & 23 13 44.00  & $+$61 26 57.8  & 30.67 (0.25) & 28.32 (0.14) &  170.8\phantom{0} (0.7\phantom{0}) & 541.2 (2.1) & IRS\,11 \\
 J23134393+6126503 & 23 13 43.93  & $+$61 26 50.3  & 3.05 (0.02)     & 27.64 (0.13) & 113.5\phantom{0} (0.4\phantom{0}) &  237.8 (1.1) & IRS\,11\,S, H$_2$O maser  \\
 J23134363+6126086 & 23 13 43.63 & $+$61 27 08.6 &  1.73 (0.04)  & 5.94 (0.04)  &  8.10 (0.09)  &  25.9 (0.4) & \\
 J23134302+6126555 & 23 13 43.02  & $+$61 26 55.5 &14.70 (0.08)      &   14.07 (0.07)   &   109.9\phantom{0} (0.5\phantom{0})   & 401.6 (1.6)  &  \\
 J23134293+6127035 & 23 13 42.93  & $+$61 27 03.5   & 1.03 (0.01)     &   4.84 (0.03)  &  14.4\phantom{0} (0.1\phantom{0})  &  37.8 (0.4)  & \\
 J23134160+6126427 & 23 13 41.60 & $+$61 26 42.7  &  1.91 (0.01) & 2.06 (0.02) & 2.73 (0.04) & 3.9 (0.3) &  \\
J23134148+6126495 & 23 13 41.48 & $+$61 26 49.5   &   4.61 (0.03)  &  5.41 (0.04)  &  6.77 (0.07)  &  8.5 (0.3) &  field star  
 \enddata
\end{deluxetable}

\end{document}